\documentclass[aps,prl,twocolumn,superscriptaddress,floatfix]{revtex4-2}

\usepackage{fontspec}
\usepackage{amsmath,amssymb,amsthm}
\usepackage{algorithm}
\usepackage{algorithmic}
\usepackage{float}
\usepackage{xcolor}
\usepackage{graphicx}
\usepackage[hidelinks]{hyperref}
\begin{document}
	
	\title{Continuous Time Quantum Walk Propagation for Irregular Temporal Graph Forecasting}
	
	\author{Jiaqi Sun}
	\author{Tianhao Li}
	\author{Zhihao Bian}
	\email{zhihaobian@jiangnan.edu.cn}
	
	\affiliation{School of Optoelectronic Information and Physical Science, Jiangnan University, Wuxi 214122, China}
	\affiliation{Wuxi Key Laboratory of Optoelectronic Intelligent Perception, Wuxi, China}
	
	\date{\today}
	
	%--------------------------------------------------------------------
\begin{abstract}
	Continuous time quantum walks (CTQWs) built on graph Laplacians can produce nonmonotonic propagation through quantum interference, providing a physically motivated propagation scheme for temporal graph signal modeling.
	We propose Quantum Walk Temporal Architecture (QWTA), which replaces classical graph propagation with a CTQW-type parameterized spectral propagator.
	QWTA embeds effective observation intervals into the spectral phase modulation of the propagation kernel.
	QWTA-Base preserves exact CTQW spectral evolution and serves as a physically faithful propagation reference.
	QWTA-GR further introduces phase soft clipping and gated residual fusion to bound the spectral phase and learn separate weights for the propagated and residual paths, respectively.
	The results show that explicit phase encoding of irregular time intervals, when combined with suitable stabilization, provides a physically motivated and competitive graph propagation design for temporal graph forecasting with missing historical observations.
\end{abstract}

	\maketitle
	
%====================================================================
\section{Introduction}
\label{sec:introduction}

Irregular time series commonly appear in real urban traffic sensing scenarios. In the running process of road networks, issues including faulty sensor hardware, communication breakdowns and road construction work may lead to missing historical measurement values. Data that should have been sampled at uniform time intervals thus becomes sparse sequences with uneven time gaps between records. Example from the Los Angeles freeway network~\cite{METR_LA}: a large number of sensors continuously collect vehicle speed data at fixed time intervals, forming a typical large-scale time series signal. However, due to unavoidable observation gaps in real deployment environments, traffic forecasting models are better framed as time series extrapolation under missing historical observations. Consequently, the model must extract spatiotemporal dependencies from historical observations that are irregularly spaced when both the number and location of effective observation time steps are uncertain.

Current graph neural network (GNN) methods typically handle missing observations via interpolation~\cite{value_completion}, masking mechanisms~\cite{Missing_Values}, or explicit temporal features~\cite{Sampled_Time_Series}. Their core message passing layers, such as GAT~\cite{GAT}, GCN~\cite{GCN}, GIN~\cite{GIN}, and Transformer~\cite{Transformer}, can accept auxiliary features like timestamps or time intervals, but they generally cannot directly embed the actual time gaps between adjacent valid observations into the parameterization of the propagation operator. Consequently, these approaches rely mainly on downstream learnable mappings to handle irregular time intervals, rather than explicitly modeling them within the graph propagation kernel.

Continuous time quantum walk (CTQW) is a fundamental dynamical process in quantum information science~\cite{CTQW_define,Origin_CTQW,QW_B}. The quantum superposition and interference exhibited during its evolution display nonmonotonic features that are absent in classical random walks, leading to many novel properties and experimental realizations in diverse systems, including photonic quantum systems~\cite{photonic_quantum_systems1,photonic_quantum_systems2,photonic_quantum_systems3}, waveguide arrays~\cite{Waveguide_QW_1,Waveguide_QW_2,Waveguide_QW_3}, cold atoms~\cite{Quantum_Walk_atoms1,Quantum_Walk_atoms2}, integrated photonic chips~\cite{Correlated_Photons} and ion traps~\cite{Quantum_Walk_Ion_Traps}.

There are already works that incorporate quantum walks into GNN algorithms~\cite{QWNN,QWGN,QWGCN}. The CTQW-type spectral propagation operator provides a means to encode time intervals directly into the propagation kernel phase modulation. As the effective observation intervals vary, different Laplace spectral components accumulate distinct phases, so the propagation output carries spectral dynamics related to the time intervals. This mechanism enables the model to utilize the actual gaps between adjacent valid observations without requiring additional interpolation or reconstruction of missing observations.

In this article, we propose a quantum walk-inspired temporal graph prediction framework, named Quantum Walk Temporal Architecture (QWTA). We replace the classical graph propagation layer with a CTQW-type parameterized spectral propagation operator, directly embedding the effective observation interval into the spectral phase factor of the propagation kernel. This yields a GNN model for temporal prediction. As the base model, QWTA-Base serves as a shared recursive spectral propagation backbone. Building on the QWTA-Base model, we apply phase soft truncation to improve stability and gate-enabled residual fusion to preserve the input paths. These modifications yield the optimized QWTA-GR model. On the METR-LA dataset~\cite{METR_LA}, we construct several controlled histories missing settings, spanning prediction scenarios from full historical observation to highly sparse histories, and we compare under a shared backbone against a variety of classical graph propagation baselines. Experimental results show that QWTA-GR achieves validation errors better than the classical propagation baselines under multiple missingness settings, and maintains competitive predictive performance in highly sparse scenarios.

The remainder of this article is organized as follows.
Sec.~\ref{sec:background} reviews GNN propagation, CTQW theory and the mapping from CTQW evolution to graph propagation operators.
Sec.~\ref{sec:model} presents the QWTA framework.
Sec.~\ref{subse:QWTA experiments} reports the experimental settings and results. Sec.~\ref{ssec:quantum_circuit} provides an initial proposal for realizing the QWTA model via quantum circuits. Finally, the concluding
remarks are given in Sec.~\ref{sec:discussion_conclusion}.

%====================================================================
\section{Theoretical Background}
\label{sec:background}
%====================================================================

\subsection{Graph propagation in GNNs}

GNNs rely on aggregating a node’s own features with its neighbors’ features under a graph structure constraint, and deriving new node representations via a learnable mapping. A graph can be defined as $G=(\mathcal{V},\mathcal{E})$, where $\mathcal{V}$ is the node set, each node represents an entity. $\mathcal{E}$ is the edge set, each edge connects two nodes and represents a relationship between entities. $|\mathcal{V}|=N$ is the number of nodes. The adjacency matrix of a graph $A\in\mathbb{R}^{N\times N}$ is an $N\times N$ matrix, with element $A_{i j} = 1$ if vertices $i$ and $j$ are adjacent, otherwise $A_{i j} = 0$. The number of edges connected to a vertex is the degree of the vertex, $\delta_i = \sum_j A_{i j}$, and the degree matrix is $\mathcal{D}=\mathrm{diag}(\delta_1,\dots,\delta_N)$.

GNNs update node features through a learnable propagation operator. Let the input node feature matrix be $\mathbf{X}\in\mathbb{R}^{N\times F}$, where $F$ denotes the input feature dimension. The output node feature matrix after a single graph propagation can be abstractly represented as

\begin{equation}
		\mathbf{X}' =
		\sigma\left(
		\mathcal{P}
		\left(
		A,\mathbf{X};\Theta
		\right)
		\right),
	\label{eq:gnn_general_prop}
\end{equation}
where
$\mathbf{X}' \in \mathbb{R}^{N\times F'}$
with $F'$ is the output feature dimension and
$\mathcal{P}(\cdot)$
denotes the graph propagation mapping.
This mapping aggregates neighborhood information and transforms features from the input node features
$\mathbf{X}$
according to the graph structure $A$.
The symbol
$\Theta$
denotes learnable parameters in the propagation mapping, and
$\sigma(\cdot)$
is a nonlinear activation function.

Eq.~\eqref{eq:gnn_general_prop}
shows that a GNN propagation layer comprises two basic steps: information is transmitted and aggregated among nodes under the influence of the adjacency matrix $A$; on this basis, the learnable parameters $\Theta$ transform the aggregated features.
Different GNN models mainly differ in how the propagation mapping
$\mathcal{P}$
uses the graph structure to define neighborhood aggregation weights, and in how
$\Theta$
participates in node feature updates. From this perspective, classical GNN propagation layers can be viewed as learnable message passing operators defined on a graph.
However, the propagation mapping
$\mathcal{P}$
is usually determined by adjacency, node features and learnable parameters.
It does not directly include the spectral phase factors associated with continuous time evolution.
To write an irregular observation interval
$\Delta t_k$
directly into the propagation kernel, one needs a graph propagation operator that varies with the time interval.

\subsection{CTQW on a graph}

Implemented based on graph topological structures and quantum coherent evolution mechanisms, CTQWs on graphs construct quantum dynamical systems purely through geometric topological properties of graphs without adopting the coin degree of freedom required by discrete time quantum walks. Specifically, the vertices of the target graph are employed as orthogonal basis vectors to span the corresponding Hilbert space. The Hamiltonian is formulated via the graph adjacency matrix or Laplacian matrix, which fully encodes the vertex connections and topological characteristics of the graph, and further determines the transition rules and dynamical evolution of quantum walkers.
Here, we use the symmetric normalized Laplacian matrix as the Hamiltonian. The form is specifically defined as
\begin{equation}
	L
	=
	I-\mathcal{D}^{-1/2}A\mathcal{D}^{-1/2}.
	\label{eq:laplacian}
\end{equation}
For an undirected graph,
$L$ is a real symmetric positive semidefinite matrix.
It therefore admits an orthogonal eigendecomposition
\begin{equation}
	L
	=
	V\Lambda V^\top,
	\label{eq:laplacian_eigendecomp}
\end{equation}
where $V=[\mathbf v_1,\ldots,\mathbf v_N]$ is an orthonormal eigenvector matrix and $\top$ denotes transpose. Its $n$th column $\mathbf v_n$ is a unit eigenvector of $L$, satisfying $L\mathbf v_n=\lambda_n\mathbf v_n$. The diagonal matrix $\Lambda=\mathrm{diag}(\lambda_1,\ldots,\lambda_N)$ contains the eigenvalues. The spectrum of the normalized Laplacian satisfies $\lambda_n\in[0,2]$. The multiplicity of the zero eigenvalue equals the number of connected components. This bounded spectral range provides a controlled frequency scale for subsequent spectral phase modulation~\cite{Graph_eigenvalues,Spectral_Graph_Theory}.

When $L$ is adopted as the Hamiltonian, the evolution of the CTQW is governed by the unitary operator
\begin{equation}
	U(\tau)
	=
	e^{-iL\tau},
	\label{eq:evolution}
\end{equation}
where $\tau$ denotes the evolution time.
If the system starts from the initial quantum state
$|\psi(0)\rangle\in\mathbb{C}^{N}$,
the evolved state becomes
\begin{equation}
	|\psi(\tau)\rangle
	=
	U(\tau)|\psi(0)\rangle.
	\label{eq:ctqw_state}
\end{equation}
% T02: unused node-probability definition removed.

Since $U(\tau)$ is a matrix function of Laplacian matrix $L$, they share the same eigenvectors $V$, and the evolution operator in Eq.~\eqref{eq:evolution} possesses an exact spectral decomposition, that is
\begin{equation}
	\begin{aligned}
		U(\tau)
		&=
		V e^{-i\Lambda\tau} V^\top
		\\
		&=
		V\mathrm{diag}
		\left(
		e^{-i\lambda_1\tau},
		\dots,
		e^{-i\lambda_N\tau}
		\right)V^\top.
	\end{aligned}
	\label{eq:spectral}
\end{equation}
The corresponding eigenvalue of
$U(\tau)$ is
$e^{-i\lambda_n\tau}$.

From a spectral perspective,
$V^\top$ projects a node domain signal into the Laplacian eigenmode space.
The diagonal matrix
$\mathrm{diag}(e^{-i\lambda_n\tau})$
applies a phase rotation to each spectral component.
The matrix $V$ then maps the modulated spectral representation back to the node domain.
Low-frequency modes have smaller $\lambda_n$, so their phases vary more slowly with $\tau$,
whereas high-frequency modes have larger $\lambda_n$, causing their phases to vary more rapidly. Consequently,
CTQW propagation is not simply a real-valued smoothing diffusion; it is determined by the superposition of complex phases across spectral modes.

Generally speaking, classical random walks use probability distributions as their basic objects, and their propagation typically diffuses toward a stationary state.
In contrast, a CTQW evolves complex probability amplitudes.
Amplitudes associated with different paths or spectral modes can interfere constructively or destructively~\cite{Origin_CTQW,Quantum_Interference,Quantum_Random_Walks}.
Quantum interference, therefore, enables a CTQW to exhibit nonmonotonic propagation responses.
This provides a physical basis for neural network operators that differ from classical diffusion-type graph propagation layers.

\subsection{Mapping CTQW to GNN propagation operators}

To employ CTQW as a propagation module in GNNs, it is necessary to generalize the evolution form of quantum states from node space to multidimensional node features.
Let the input node feature matrix for a single layer of graph propagation be
$\mathbf{X}\in\mathbb{R}^{N\times F}$,
where $F$ is the feature dimension per node. Denote the feature vector of node $n$ by
$\mathbf{x}_n\in\mathbb{R}^{F}$, then the input feature matrix can be written as
\begin{equation}
	\mathbf{X}
	=
	\begin{bmatrix}
		\mathbf{x}_1^\top \\
		\mathbf{x}_2^\top \\
		\vdots \\
		\mathbf{x}_N^\top
	\end{bmatrix}
	=
	\left[
	\mathbf{x}_1,
	\mathbf{x}_2,
	\dots,
	\mathbf{x}_N
	\right]^\top
	\in
	\mathbb{R}^{N\times F},
	\label{eq:node_feature_matrix}
\end{equation}
here
$\mathbf{x}_n$
denotes the feature vector corresponding to the $n$th node.
Thus,
$\mathbf{X}$
is not merely a set of features, but a multichannel graph signal defined on node positions.
The row index of the matrix corresponds to node positions, and the column index corresponds to node feature channels.
From the perspective of the representation space,
$\mathbf{X}$ can be understood as a coupled representation between the node position space and the feature space.
In other words,
$\mathbf{X}$
simultaneously contains information about node locations and node features and can be viewed as a position–feature coupling matrix.
The term coupling here denotes the binding relationship between node positions and node features in GNNs, and should not be interpreted as an entangled state in the strict sense of quantum mechanics.

In standard CTQW, the evolution operator
$U(\tau)$
acts on quantum states in node space.
To apply it to multidimensional node feature propagation, we embed the real-valued feature matrix
$\mathbf{X}$
into the complex domain with zero imaginary part, and applies the CTQW evolution operator to the node dimension
\begin{equation}
	\mathbf{X}'
	=
	U(\tau)\mathbf{X},
	\label{eq:ctqw_feature_prop}
\end{equation}
where
$\mathbf{X}'$
denotes the output feature matrix after CTQW-type propagation.
Here, the rows and columns of
$U(\tau)$
corresponds to propagation in the node position space. Therefore, Eq.~\eqref{eq:ctqw_feature_prop} describes the propagation of CTQW evolution in the node position space, wherein the positional information of the nodes and the multichannel feature information they carry are coupled and change simultaneously.

%====================================================================
\section{QWTA Model}
\label{sec:model}
\begin{figure*}[t]
	\centering
	\includegraphics[width=\textwidth]{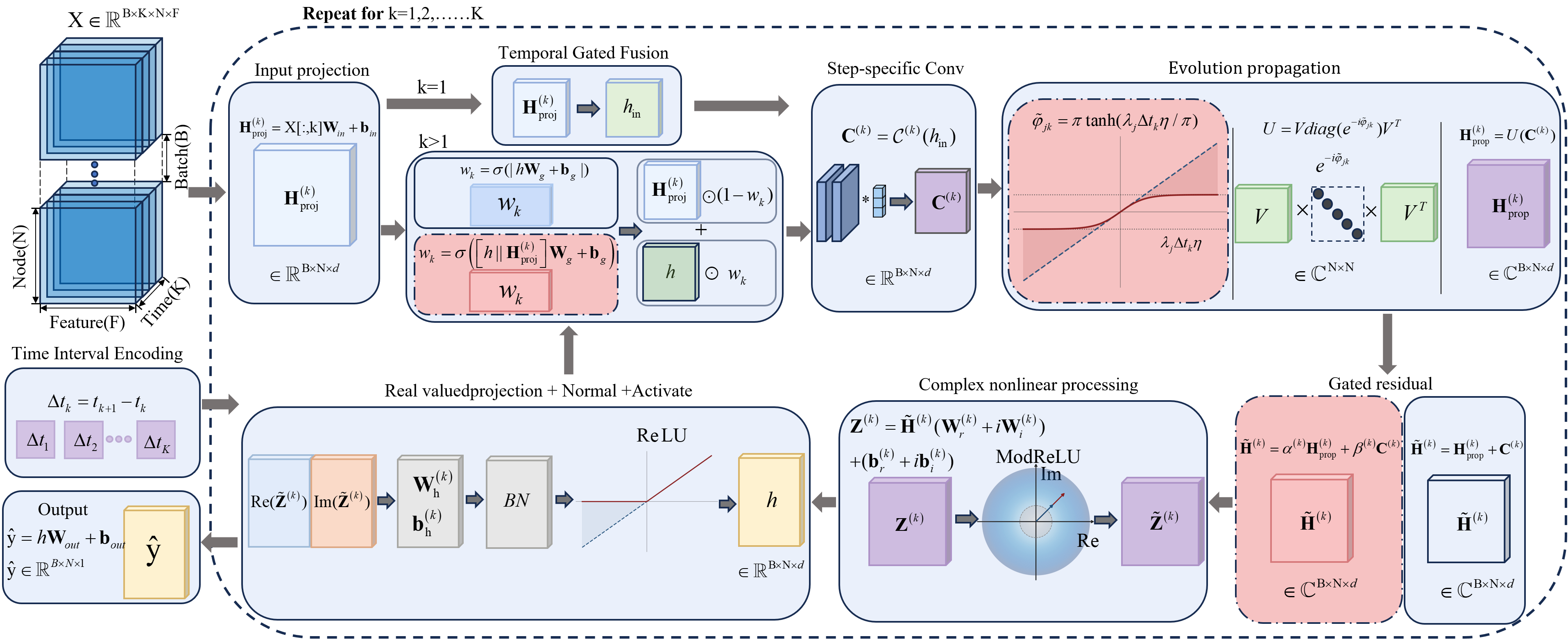}
	\caption{Forward propagation pipeline of QWTA. The model input is a batch tensor of irregular historical observations,
		$\mathbf{X}\in\mathbb{R}^{B\times K\times N\times F}$. The input passes through input projection, recurrent hidden fusion, CTQW-type spectral propagation, and complex-valued nonlinear processing to generate the recurrent hidden representation.
		 The red dashed boxes mark QWTA-GR-specific operations. The final hidden representation is projected into the real domain to generate the prediction.}
	\label{fig:model}
\end{figure*}

%====================================================================
This section describes the model architecture in accordance with the actual data flow of QWTA (Fig.~\ref{fig:model}). First, we perform temporal information preprocessing, feature encoding, and historical information reconstruction on the original irregular time series graph signals. Next, the time interval driven quantum evolution propagation is carried out using a CTQW-type spectral propagation operator. Finally, by applying a complex-valued nonlinear activation, real-valued projection, and an output layer, the predictive results are obtained. Building on this, the article further presents the QWTA-GR expansion module.

\subsection{Raw information preprocessing and encoding}
\subsubsection{1. Temporal information preprocessing}

In a traffic network graph, the observations of node $n$ on the time sequence $\{t_1,\dots,t_K\}$ form a set of time series signals, where the feature information of node $n$ is represented as $\mathbf{x}_n(t_k)\in\mathbb{R}^F$, containing observed features such as velocity and a set of timestamps. Typically, observations are sampled at equal intervals. However, if sampling gaps occur, the intervals between adjacent time points are not necessarily the same, leading to irregular time series signals. To simulate irregular timing, we represent irregular time series by missing time points.

\begin{figure}[htbp]
	\centering
	\includegraphics[width=\linewidth]{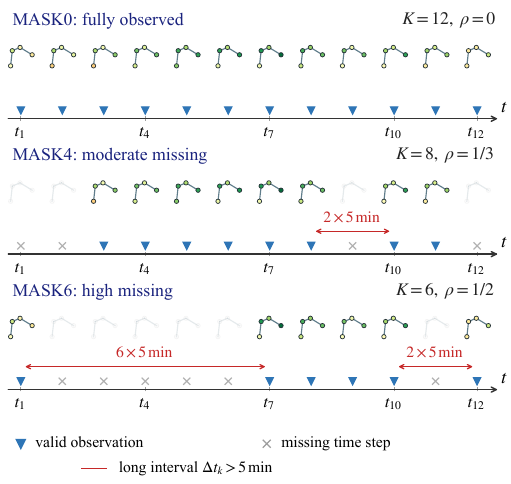}
	\caption{Temporal graph snapshot sequences under different masking settings. The figure shows traffic graph observation sequences for three representative MASK configurations, corresponding to complete observation, moderate missingness and high missingness. Valid observation steps are shown as clear graph snapshots, whereas missing steps are faded. For a given masking scheme, the missing measurements occur randomly. In this schematic, $t_j$ denotes original sample $j$ and corresponds to model timestamp $j-1$.}
	\label{fig:irregular_graph_snapshots}
\end{figure}

Represent the observation results of network node $n$ at different times using the set $\bigl\{(t_k,\mathbf{x}_n(t_k))\bigr\}_{k=1}^{K}$, where the time intervals between adjacent time nodes satisfy $\Delta t_k=t_{k+1}-t_k$ for $k<K$ (as shown in Fig.~\ref{fig:irregular_graph_snapshots}). The corresponding missing rate $\rho$ is defined as

\begin{equation}
	\rho = \frac{T_{\mathrm{full}} - K}{T_{\mathrm{full}}}.
\end{equation}
We define the original total number of time steps as $T_{\mathrm{full}} = 12$ and the number of valid observation steps retained after masking as $K$. Here $j$ indexes positions in the complete sequence starting from 1, whereas $k$ in the algorithms indexes retained observations. If the $k$th retained observation is at position $j_k$ in the complete sequence, its model timestamp, measured in units of the original sampling interval, is $t_k=j_k-1$.
For example, MASK0 corresponds to $K=12$ and $\rho=0$;
MASK4 corresponds to $K=8$ and $\rho=1/3$;
MASK6 corresponds to $K=6$ and $\rho=1/2$.
In Fig.~\ref{fig:irregular_graph_snapshots}, consecutive observations are 5 minutes apart in the fully observed sequence. The highlighted gap in the moderate missingness example spans two sampling intervals because one observation is missing, giving \(2\times5=10\) minutes. In the high missingness example, five consecutive missing observations make the gap span six sampling intervals, giving \(6\times5=30\) minutes. Note that under high missingness, missing observations may occur in multiple disjoint segments, producing several longer intervals between observations within the same sequence.

To evaluate the model’s performance under different levels of missing data, we design 11 controlled masking schemes, denoted MASK0--MASK10, corresponding to valid observation counts
$K\in\{12,11,\dots,2\}$, covering the evaluation range from complete historical observations to highly missing historical observations. Note that for a given masking scheme, the missing measurements occur randomly, i.e., the elements of the original uniform temporal interval observed data set are randomly omitted.
%Given the first $K$ irregular observations, the goal is to predict the node states at the next time step, $\hat{\mathbf{y}}\in\mathbb{R}^{B\times N}$.

At irregular time intervals, the form of the evolution operator can be written as
\begin{equation}
	U(\Delta t_k) = V\,\mathrm{diag}\!\left(
	e^{-i\lambda_1\Delta t_k},\dots,
	e^{-i\lambda_N\Delta t_k}
	\right)V^\top.
	\label{eq:phase_encoding}
\end{equation}
Propagation intervals are computed from the retained timestamps and the fixed target time and used in the spectral phase calculation. As $\Delta t_k$ changes, the phase modulation corresponding to different spectral components also changes accordingly, thereby providing the propagation representation that depends on these intervals.

\subsubsection{2. Encoding the feature matrix}

For the sake of facilitating computer learning, we perform batch processing on the input dataset.
Let the input tensor be
$\mathbf{X}\in\mathbb{R}^{B\times K\times N\times F}$,
where $B$ is the batch size,
$K$ is the number of valid observation time steps,
$N$ is the number of nodes in the traffic network and
$F$ is the input feature dimension.
For the $k$th valid observation time step, the input slice is defined as $\mathbf{X}^{(k)}
	=
	\mathbf{X}[:,k,:,:]
	\in
	\mathbb{R}^{B\times N\times F}$.
This projection encodes each node's speed and timestamp features in the same $d$-dimensional space for recurrent fusion and spectral propagation.
For QWTA-Base, a shared input affine map projects the current node features into a $d$-dimensional hidden space:
\begin{equation}
	\begin{aligned}
		\mathbf{H}_{\mathrm{proj}}^{(k)}
		&=
		\mathbf{X}^{(k)}\mathbf{W}_{\mathrm{in}}
		+
		\mathbf{b}_{\mathrm{in}},
	\end{aligned}
	\label{eq:input_proj}
\end{equation}
where
$\mathbf{W}_{\mathrm{in}} \in \mathbb{R}^{F\times d}$
is the input feature projection matrix, used to map the $F$-dimensional raw features of each node at the current valid observation time to the $d$-dimensional hidden representation space.
The term $\mathbf{b}_{\mathrm{in}} \in \mathbb{R}^{d}$ is the corresponding learnable bias and provides an affine offset in the linear mapping.
Its weight and bias are shared across valid observation steps; graph propagation is performed by the subsequent CTQW operator.
Therefore,
$\mathbf{H}_{\mathrm{proj}}^{(k)} \in \mathbb{R}^{B\times N\times d}$
denotes the node hidden feature matrix after input encoding for the $k$th valid observation time.

\subsubsection{3. Historical information reconstruction}

The input projection above provides a hidden representation of the current valid observation, while historical information is retained in the hidden state output by the preceding valid observation step. From the second valid observation step onward, the model fuses these two representations to construct the input to spectral propagation at the current step.

The loop variables $h$, $h_{\mathrm{in}}$ and $w_k$ are updated in place. At the beginning of step $k>1$, $h$ stores the real-valued recurrent output of step $k-1$. For QWTA-Base, $w_k$ is computed from this state at step $k>1$, immediately before input fusion. The fused state $h_{\mathrm{in}}$ combines this recurrent output with the current observation projection. After the block is evaluated, $h$ is updated.

For the $k$th valid observation time, the input projection layer first generates the hidden representation of the current observation
$\mathbf{H}_{\mathrm{proj}}^{(k)}$.
To enable subsequent spectral propagation to utilize both the current observation information and the recursive state preserved from previous valid observation steps, we construct an input hidden representation
$h_{\mathrm{in}}\in\mathbb{R}^{B\times N\times d}$ prior to spectral propagation.

For the first valid observation time, since there is no preceding recursive hidden representation, the propagation input is directly taken as the projection of the current observation
$h_{\mathrm{in}}
	=
	\mathbf{H}_{\mathrm{proj}}^{(1)}$. For $k>1$, the loop variable
$h\in\mathbb{R}^{B\times N\times d}$
stores the recurrent hidden representation output from the previous effective observation step.
For QWTA-Base, the gate used at step $k>1$ is computed from the preceding recurrent output with a shared affine map:
\begin{equation}
 w_k=\sigma\!\left(\left|h\mathbf W_g+\mathbf b_g\right|\right).
 \label{eq:temporal_gate_weight}
\end{equation}
Here $\mathbf W_g\in\mathbb R^{d\times1}$ and $\mathbf b_g\in\mathbb R$ are shared across steps. The absolute value is applied elementwise before the sigmoid~\cite{Sigmod}. The resulting gate $w_k\in\mathbb R^{B\times N\times1}$ is broadcast over the hidden feature channels in the fusion below.

Subsequently, the propagation input at the current time step is obtained by adaptively fusing the existing recursive hidden representation with the current observation projection,
\begin{equation}
		h_{\mathrm{in}}
		=
		w_k\odot h
		+
		\left(1-w_k\right)
		\odot
		\mathbf{H}_{\mathrm{proj}}^{(k)},
		\qquad k>1,
	\label{eq:temporal_gated_input}
\end{equation}
where the symbol $\odot$ denotes elementwise Hadamard multiplication.
When $w_k$ is large, the propagated input places greater emphasis on previously recurrent hidden representations.
When $w_k$ is small, the propagated input relies more strongly on the current valid observation features.
Therefore,
$h_{\mathrm{in}} \in \mathbb{R}^{B\times N\times d}$
represents the hidden representation after fusing historical propagation information with the current observation information in the current valid observation step, and serves as the actual input to the subsequent CTQW-type spectral propagation operators.

\subsection{Quantum evolution process}

QWTA-Base uses the normalized Laplacian $L$ as the Hamiltonian for exact CTQW spectral propagation. Before propagation, the fused real-valued state passes through a step-specific convolution layer (Conv2d). The layer takes a $B\times d\times1\times N$ tensor, with kernel size $(3,1)$ and padding $(1,0)$. Returning to the node feature layout gives
\begin{equation}
 \mathbf C^{(k)}=\mathcal C^{(k)}(h_{\mathrm{in}})
 \in\mathbb R^{B\times N\times d}.
 \label{eq:preprop_conv}
\end{equation}
Here $\mathcal C^{(k)}$ denotes the Conv2d mapping with its learnable kernel and bias. The convolution acts on the singleton temporal dimension of the current step; recurrence carries information across observations. The propagation output is
\begin{equation}
	\begin{aligned}
		\mathbf{H}^{(k)}_{\mathrm{prop}}
		&= U(\Delta t_k)\mathbf C^{(k)}
		\\
		&= V\,\mathrm{diag}\!\left(
		e^{-i\varphi_{1k}},\dots,
		e^{-i\varphi_{Nk}}
		\right)V^\top \mathbf C^{(k)},
	\end{aligned}
	\label{eq:quantum_prop}
\end{equation}
where the phase is defined as $\varphi_{nk}:=\lambda_n\Delta t_k\eta$ and the complex-valued propagation feature $\mathbf{H}^{(k)}_{\mathrm{prop}} \in \mathbb{C}^{B\times N\times d}$. \(\Delta t_k\) denotes the interval to the next valid observation, or to the fixed prediction target at the last step (Sec.~III.A). \(\eta\in\mathbb R\) is a learnable global time scaling coefficient. Accordingly, the effective evolution parameter within the propagation operator is \(\eta\Delta t_k\).

The introduction of \(\eta\) aims to construct a learnable mapping between the observation time scale and the spectral propagation time scale. Although \(\Delta t_k\) is determined by observation timestamps, directly adopting this interval as the evolution parameter of the CTQW-type operator does not necessarily yield propagation features suitable for prediction tasks. During training, the model optimizes \(\eta\) to adjust the phase accumulation rate of each spectral mode, so that the propagation scale adapts to the spatiotemporal dependencies embedded in data. Within one model, \(\eta\) is shared across all samples and valid observation steps, while each step still uses its corresponding \(\Delta t_k\). In QWTA-Base, $\eta$ is a single signed scalar; the phase uses it directly, without an absolute value transformation or phase clipping. Hence, the relative relations among the original irregular intervals are preserved. After training, \(\eta\) is fixed to its learned value and remains unchanged during inference. The evolution parameter for each prediction is jointly determined by this fixed coefficient and the actual observation intervals of the current input. In this way, MASK schemes define observation intervals, and \(\eta\) modulates the scale of these intervals inside the propagation operator. The two quantities carry distinct and complementary meanings.

\subsection{Nonlinear activation and output layer}
%\subsubsection{Nonlinear activation process}
General transportation networks contain a large number of nodes, and deep propagation in graphs often leads to feature attenuation or gradient vanishing. To enhance the stability and learning performance of the model, we introduce a residual structure to stabilize information transmission across multiple layers and alleviate these issues.
The propagated features are combined with the post-convolution features $\mathbf C^{(k)}$ through a fixed residual path:
\begin{equation}
		\widetilde{\mathbf{H}}^{(k)}
		=
		\mathbf{H}_{\mathrm{prop}}^{(k)}
		+
		\mathbf C^{(k)}.
	\label{eq:base_residual}
\end{equation}

Although the complex features obtained from residual fusion preserve the propagation information and pre-propagation input information, there remains a lack of learnable complex transforms among different hidden feature channels. To further leverage the complex representations obtained from quantum propagation, we introduce a complex linear layer that enables learnable scaling, rotation, and channel mixing of the hidden features, allowing different feature components to be recombined in the complex domain and thereby enhance feature expressive power. The feature information matrix is transformed by a ComplexLinear layer,

\begin{equation}
		\mathbf{Z}^{(k)}
		= \tilde{\mathbf{H}}^{(k)}\,(\mathbf{W}^{(k)}_r + i\mathbf{W}^{(k)}_i)
		+ (\mathbf{b}^{(k)}_r + i\mathbf{b}^{(k)}_i),
	\label{eq:complex_linear}
\end{equation}
here $\mathbf{W}^{(k)}_r,\mathbf{W}^{(k)}_i\in\mathbb{R}^{d\times d}$ are the real and imaginary weight components, respectively.
The corresponding biases $\mathbf{b}^{(k)}_r,\mathbf{b}^{(k)}_i\in\mathbb{R}^{d}$ are independent learnable parameters. The residual output and complex features have shape $B\times N\times d$; the real-valued residual input is embedded in the complex domain for the sum.

The multilayer stacking of purely linear graph convolutional networks is still essentially a linear transformation, incapable of approximating arbitrary nonlinear functions, and it is difficult to characterize the complex dependencies between node features and topological structure in graph data. Nonlinear activation functions introduce nonlinear mappings after feature aggregation, breaking the constraint of linear superposition, enhancing the model’s representational capacity, and enabling the network to learn the joint patterns of high order structures and features on graphs. Apply ModReLU activation to the feature information matrix obtained from Eq.~\eqref{eq:complex_linear} yields the following
\begin{equation}
	\begin{aligned}
		\tilde{\mathbf{Z}}^{(k)} &= \mathrm{ModReLU}(\mathbf{Z}^{(k)})   \\
		&= \mathrm{ReLU}(|\mathbf{Z}^{(k)}|+\mathbf b_{\mathrm{relu}}^{(k)})\cdot e^{i\angle \mathbf{Z}^{(k)}}.
	\end{aligned}
	\label{eq:modrelu}
\end{equation}
This activation applies a nonlinear transformation to the modulus of complex features while keeping the complex direction invariant, where $\mathbf b_{\mathrm{relu}}^{(k)}\in\mathbb{R}^{d}$ is a learnable bias and $\angle(\cdot)$ denotes the complex argument. After ModReLU processing, the resulting $\tilde{\mathbf{Z}}^{(k)}$ remains a complex feature. Since subsequent feature transformations use real-valued neural networks, $\tilde{\mathbf{Z}}^{(k)}$ is decomposed into its real and imaginary parts and then concatenated. The real and imaginary parts together constitute the complete representation, so this transformation does not lose information from the aforementioned complex features. Subsequently, further feature extraction is performed through a real-valued linear mapping, BatchNorm~\cite{BN}, ReLU~\cite{ReLU}, as shown by the following

\begin{equation}
		h
		=\mathrm{ReLU}\!(\mathrm{BN}\!(
		\left[
		\mathrm{Re}(\tilde{\mathbf{Z}}^{(k)})\,\Vert\,
		\mathrm{Im}(\tilde{\mathbf{Z}}^{(k)})
		\right]
		\mathbf{W}_{\text{h}}^{(k)}
		+ \mathbf{b}_{\text{h}}^{(k)})),
	\label{eq:real_proj}
\end{equation}
where $\mathbf{W}^{(k)}_{\text{h}}\in \mathbb{R}^{2d\times d}$ and $\mathbf{b}^{(k)}_{\text{h}}\in \mathbb{R}^d$ are the learnable weight and bias. Concatenation produces a $B\times N\times2d$ real-valued tensor; the projection, BatchNorm and ReLU return $h\in\mathbb R^{B\times N\times d}$. Here $[\,\cdot\,\Vert\,\cdot\,]$ denotes concatenation along the last dimension, in the order shown, with all other dimensions unchanged.

The above steps constitute a complete QWTA block. After $K$ loop iterations, the final forecast is produced by the output linear layer,
\begin{equation}
		\hat{\mathbf{y}}
		=
		h\mathbf{W}_{\text{out}}
		+ \mathbf{b}_{\text{out}},
	\label{eq:output_layer}
\end{equation}
where
$\mathbf{W}_{\text{out}}\in\mathbb{R}^{d\times 1}$
and
$\mathbf{b}_{\text{out}}\in\mathbb{R}$ are the learnable weight and bias, respectively. Up to this point, the QWTA model forms a full prediction pipeline from spectral domain propagation to real-valued output. This fundamental process is termed QWTA-Base (see Fig.~\ref{fig:model}). The whole computation
process of QWTA-Base is represented by the pseudo code in Algorithm~\ref{alg:qwta_forward}.

This model uses normalized speed and unnormalized original timestamps for the $F=2$ input features, projects them into a hidden space of dimension $d=64$, and processes the $K$ valid observations sequentially.
In QWTA-Base, the input projection, temporal gate, global $\eta$ and output projection are shared across steps. The convolution, ComplexLinear, ModReLU, real-valued projection and BatchNorm parameters are step specific. The parameter set $\Theta_{\mathrm{Base}}$ comprises these parameters. BatchNorm uses batch statistics during training and stored running statistics during inference.
In the algorithms below, $\mathbf C^{(k)}$ denotes the output of the convolution in Eq.~\eqref{eq:preprop_conv}. Both the QWTA-Base fixed residual and the QWTA-GR self path use the post-convolution features $\mathbf C^{(k)}$; QWTA-GR additionally applies the learned residual weights.

In the algorithms, timestamp differences are computed separately for each batch sample, using the time definition in Sec.~III.A.
For $k>1$, the gate is computed from the preceding recurrent state immediately before the current input fusion. No temporal gate is needed for the first step.

\begin{algorithm}[H]
\caption{QWTA-Base Forward Pass}
\label{alg:qwta_forward}
\small
\begin{algorithmic}[1]
\STATE \textbf{Input:} $\mathbf X\in\mathbb R^{B\times K\times N\times F}$,
eigenvalues $\{\lambda_n\}_{n=1}^N$, eigenvectors $V\in\mathbb R^{N\times N}$, model parameters $\Theta_{\mathrm{Base}}$
\STATE \textbf{Precompute:} $\Delta t_k\gets t_{k+1}-t_k$ for $k<K$, from the timestamp channel; $\Delta t_K\gets12-t_K$
\STATE Initialize $h\gets\mathbf0\in\mathbb R^{B\times N\times d}$
\FOR{$k=1$ \textbf{to} $K$}
\STATE $\mathbf X^{(k)}\gets\mathbf X[:,k,:,:]$
\STATE $\mathbf H_{\mathrm{proj}}^{(k)}\gets\mathbf X^{(k)}\mathbf W_{\mathrm{in}}+\mathbf b_{\mathrm{in}}$
\IF{$k=1$}
\STATE $h_{\mathrm{in}}\gets\mathbf H_{\mathrm{proj}}^{(k)}$
\ELSE
\STATE $w_k\gets\sigma(|h\mathbf W_g+\mathbf b_g|)$
\STATE $h_{\mathrm{in}}\gets w_k\odot h+(1-w_k)\odot\mathbf H_{\mathrm{proj}}^{(k)}$
\ENDIF
\STATE $\mathbf C^{(k)}\gets\mathcal C^{(k)}(h_{\mathrm{in}})$
\STATE $\varphi_{nk}\gets\lambda_n\Delta t_k\eta$, $n=1,\ldots,N$
\STATE $U\gets V\,\mathrm{diag}(e^{-i\varphi_{1k}},\ldots,e^{-i\varphi_{Nk}})V^\top$
\STATE $\mathbf H_{\mathrm{prop}}^{(k)}\gets U\mathbf C^{(k)}$
\STATE $\widetilde{\mathbf H}^{(k)}\gets\mathbf H_{\mathrm{prop}}^{(k)}+\mathbf C^{(k)}$
\STATE $\mathbf Z^{(k)}\gets\mathrm{ComplexLinear}^{(k)}(\widetilde{\mathbf H}^{(k)})$
\STATE $\widetilde{\mathbf Z}^{(k)}\gets\mathrm{ModReLU}^{(k)}(\mathbf Z^{(k)})$
\STATE $\mathbf R^{(k)}\gets[\mathrm{Re}(\widetilde{\mathbf Z}^{(k)})\,\Vert\,\mathrm{Im}(\widetilde{\mathbf Z}^{(k)})]$
\STATE $h\gets\mathrm{ReLU}(\mathrm{BN}^{(k)}(\mathbf R^{(k)}\mathbf W_{\mathrm h}^{(k)}+\mathbf b_{\mathrm h}^{(k)}))$
\ENDFOR
\STATE $\hat{\mathbf y}\gets h\mathbf W_{\mathrm{out}}+\mathbf b_{\mathrm{out}}$
\STATE \textbf{Return:} $\hat{\mathbf y}\in\mathbb R^{B\times N\times1}$
\end{algorithmic}
\end{algorithm}

\subsection{Gated residual fusion extension: QWTA-GR}

QWTA-GR represents a structural extension built upon QWTA-Base, with improvements targeting the spectral propagation stability and residual fusion flexibility inside the QWTA block. The core propagation design has two components. First, soft truncation is applied to the spectral phase during quantum evolution propagation to mitigate rapid phase oscillations over large time intervals. Second, gated residual fusion is introduced after propagation, enabling the model to adaptively balance the relative contributions from the spectral propagation representation and the pre-propagation input representation.

Before spectral propagation, QWTA-GR combines the previous recurrent state with the current observation projection through a channelwise temporal gate. For $k>1$, $h$ denotes the recurrent output from step $k-1$, and the gate is
\begin{equation}
 w_k=\sigma\!\left(
 [h\,\Vert\,\mathbf H_{\mathrm{proj}}^{(k)}]\mathbf W_g+\mathbf b_g
 \right).
 \label{eq:gr_temporal_gate}
\end{equation}
Here $\mathbf W_g\in\mathbb R^{2d\times d}$ and $\mathbf b_g\in\mathbb R^d$ are the QWTA-GR temporal gate parameters, shared across observation steps. The input projection also uses shared $\mathbf W_{\mathrm{in}}$ and $\mathbf b_{\mathrm{in}}$. Concatenation along the hidden feature dimension gives a $B\times N\times2d$ tensor, and $w_k\in\mathbb R^{B\times N\times d}$ assigns a separate weight to each hidden channel. No absolute value is applied before the sigmoid.

When $k=1$, set $h_{\mathrm{in}}=\mathbf H_{\mathrm{proj}}^{(1)}$ directly. For $k>1$, first compute $w_k$ using the temporal gate equation above, then substitute it into Eq.~\eqref{eq:temporal_gated_input} to combine the previous recurrent state with the current observation projection and obtain $h_{\mathrm{in}}$.

For large \(\Delta t_k\), the phase term in Eq.~\eqref{eq:quantum_prop} may attain a considerable magnitude. Relative phases across distinct spectral modes evolve rapidly with the time interval, causing severe oscillations of the propagated representation and its gradients throughout training. To alleviate optimization sensitivity within large phase regimes, QWTA-GR imposes soft truncation on the raw spectral phase, namely

\begin{equation}
	\tilde{\varphi}_{nk}
	= \pi\tanh\!\left(\frac{\lambda_n\Delta t_k\eta}{\pi}\right).
	\label{eq:phase_clip}
\end{equation}

Following phase soft truncation, the quantum evolution of feature representations within QWTA-GR is written as
\begin{equation}
	\begin{aligned}
		\mathbf{H}^{(k)}_{\text{prop}}
		&= \tilde{U}(\Delta t_k)\,\mathbf C^{(k)}
		\\
		&= V\,\mathrm{diag}\!\left(
		e^{-i\tilde{\varphi}_{1k}},\dots,e^{-i\tilde{\varphi}_{Nk}}
		\right)V^\top \mathbf C^{(k)}.
	\end{aligned}
	\label{eq:H_phase_clip}
\end{equation}
For small magnitudes of the parameter $x$, $\tanh(x)\approx x$, and Eq.~\eqref{eq:phase_clip} recovers the original phase. With growing phase magnitude, soft truncation bounds the spectral phase to a finite interval, which suppresses fast phase fluctuations arising from large time intervals or large time scaling parameters.

To mitigate deep network oversmoothing and retain original features, we adopt a gated residual mechanism. Adaptive tuning of the fusion ratio between propagated information and initial input effectively boosts the model representation power. Drawing inspiration from classic ResNet and LSTM neural networks, we incorporate a gated residual to substitute Eq.~\eqref{eq:base_residual},

\begin{equation}
	\widetilde{\mathbf{H}}^{(k)}
	=
	\alpha^{(k)}\,
	\mathbf{H}_{\mathrm{prop}}^{(k)}
	+
	\beta^{(k)}\,
	\mathbf C^{(k)},
	\label{eq:Gated_residual}
\end{equation}
where $\alpha^{(k)}= \sigma(g_{\text{prop}}^{(k)})$, $\beta^{(k)}= \sigma(g_{\text{self}}^{(k)})$. The variables
$g_{\text{prop}}^{(k)}$
and
$g_{\text{self}}^{(k)}$
are learnable scalars initialized to $0$.
At initialization, $\alpha^{(k)}=\beta^{(k)}=0.5$ for every step $k$. With both coefficients initialized to 0.5, the fusion initially computes
the average of the two path outputs.
Eq.~\eqref{eq:Gated_residual} endows the model with the capability to adaptively assign mixing weights between spectrally propagated features and locally preserved features. Therefore, we refer to this extension as gated residual (GR) fusion extension (Fig.~\ref{fig:model}), rather than treating it as a dynamic distance or dynamic topology learning module.

Algorithm~\ref{alg:qwta_gr_forward} summarizes the QWTA-GR computation. The QWTA-GR parameter set comprises shared input projection, temporal gate, phase scale and output projection parameters, together with step-specific convolution, ComplexLinear, ModReLU, real-valued projection, BatchNorm and residual gate parameters.

\begin{algorithm}[H]
	\caption{QWTA-GR Forward Pass}
	\label{alg:qwta_gr_forward}
	\small
	\begin{algorithmic}[1]
		\STATE \textbf{Input:} $\mathbf X\in\mathbb R^{B\times K\times N\times F}$,
		eigenvalues $\{\lambda_n\}_{n=1}^N$, eigenvectors $V\in\mathbb R^{N\times N}$, model parameters $\Theta_{\mathrm{GR}}$
		\STATE \textbf{Precompute:} $\Delta t_k\gets t_{k+1}-t_k$ for $k<K$, from the timestamp channel; $\Delta t_K\gets12-t_K$
		\STATE Initialize $h\gets\mathbf0\in\mathbb R^{B\times N\times d}$
		\FOR{$k=1$ \textbf{to} $K$}
		\STATE $\mathbf X^{(k)}\gets\mathbf X[:,k,:,:]$
		\STATE $\mathbf H_{\mathrm{proj}}^{(k)}\gets\mathbf X^{(k)}\mathbf W_{\mathrm{in}}+\mathbf b_{\mathrm{in}}$
		\IF{$k=1$}
		\STATE $h_{\mathrm{in}}\gets\mathbf H_{\mathrm{proj}}^{(k)}$
		\ELSE
		\STATE $w_k\gets\sigma\!\left([h\,\Vert\,\mathbf H_{\mathrm{proj}}^{(k)}]\mathbf W_g+\mathbf b_g\right)$
		\STATE $h_{\mathrm{in}}\gets w_k\odot h+(1-w_k)\odot\mathbf H_{\mathrm{proj}}^{(k)}$
		\ENDIF
		\STATE $\mathbf C^{(k)}\gets\mathcal C^{(k)}(h_{\mathrm{in}})$
		\STATE $\varphi_{nk}\gets\lambda_n\Delta t_k\eta$, $n=1,\ldots,N$
		\STATE $\widetilde\varphi_{nk}\gets\pi\tanh(\varphi_{nk}/\pi)$, $n=1,\ldots,N$
		\STATE $U\gets V\,\mathrm{diag}(e^{-i\widetilde\varphi_{1k}},\ldots,e^{-i\widetilde\varphi_{Nk}})V^\top$
		\STATE $\mathbf H_{\mathrm{prop}}^{(k)}\gets U\mathbf C^{(k)}$
		\STATE $\alpha^{(k)}\gets\sigma(g_{\mathrm{prop}}^{(k)})$, $\beta^{(k)}\gets\sigma(g_{\mathrm{self}}^{(k)})$
		\STATE $\widetilde{\mathbf H}^{(k)}\gets\alpha^{(k)}\mathbf H_{\mathrm{prop}}^{(k)}+\beta^{(k)}\mathbf C^{(k)}$
		\STATE $\mathbf Z^{(k)}\gets\mathrm{ComplexLinear}^{(k)}(\widetilde{\mathbf H}^{(k)})$
		\STATE $\widetilde{\mathbf Z}^{(k)}\gets\mathrm{ModReLU}^{(k)}(\mathbf Z^{(k)})$
		\STATE $\mathbf R^{(k)}\gets[\mathrm{Re}(\widetilde{\mathbf Z}^{(k)})\,\Vert\,\mathrm{Im}(\widetilde{\mathbf Z}^{(k)})]$
		\STATE $h\gets\mathrm{ReLU}(\mathrm{BN}^{(k)}(\mathbf R^{(k)}\mathbf W_{\mathrm h}^{(k)}+\mathbf b_{\mathrm h}^{(k)}))$
		\ENDFOR
		\STATE $\hat{\mathbf y}\gets h\mathbf W_{\mathrm{out}}+\mathbf b_{\mathrm{out}}$
		\STATE \textbf{Return:} $\hat{\mathbf y}\in\mathbb R^{B\times N\times1}$
	\end{algorithmic}
\end{algorithm}

\flushbottom

%====================================================================
\section{Experiments and Results}
\label{subse:QWTA experiments}
%====================================================================

\subsection{Training and experimental settings}

In this article we evaluate the models on the METR-LA traffic speed dataset~\cite{METR_LA}. This dataset contains four months of speed observations recorded at 5-minute intervals by 207 loop detectors deployed across the Los Angeles highway network. Following the standard protocol, we split the dataset chronologically into training, validation and test sets with a 7:2:1 ratio.

METR-LA provides traffic speed observations at regular intervals. We apply controlled masks within each 12-step historical input window to construct observation sequences at different missing rates. Timestamp processing follows the convention in Sec.~III.A: retained observations keep their original indices, the forecast target is fixed at index 12, and the final propagation interval is \(12-t_K\), where \(t_K\) is the last retained observation index.

We compare QWTA with four classical baselines: GAT~\cite{GAT}, GCN~\cite{GCN}, GIN~\cite{GIN} and Transformer~\cite{Transformer}. We adapt the baselines to irregular time series using a stepwise recurrent framework. All models use the same chronological data split, receive normalized speed and the timestamps of retained observations as inputs, and use validation MSE as the primary evaluation metric. QWTA explicitly uses \(\Delta t_k\) to parameterize its propagation operator.

To facilitate reproducibility, we record the random seed used in each run. A random seed is the initialization value for a pseudorandom number generator and determines the sequence of pseudorandom numbers it generates. For each model and MASK setting in these comparison experiments, we perform one run with each of the random seeds 40, 41, and 42, giving three runs in total.

All models are trained with mean squared error (MSE) as the loss function for the one-step forecasting task. Experiments are conducted on a single GPU with a batch size of 256. The AdamW optimizer~\cite{AdamW_TPAMI} is adopted together with a cosine annealing learning rate scheduler~\cite{SGDR}. The source code for QWTA is available on GitHub~\cite{QWTA_code}.

\subsection{Main comparison between classical baselines and QWTA models}
\label{ssec:main_results}
\label{ssec:extended_comparison}

\begin{table*}[t]
	\caption{
		Validation MSE on the METR-LA dataset for four classical graph propagation baselines (GAT, GCN, Transformer and GIN), QWTA-Base and QWTA-GR under MASK0--MASK10.
		For each model and MASK setting, we conduct three runs with random seeds 40, 41, and 42. We record the lowest validation MSE from each run and report the mean \(\pm\) sample standard deviation across the three runs.
		Bold indicates the lowest mean among all models in that row.
		Underlining indicates the lowest mean among the four classical baselines in that row, namely the per-setting Best Classical reference.
		Ties are underlined simultaneously.
		In the macro average row, the underline indicates the strongest single classical baseline, GCN.
	}
	\label{tab:main_results}
	\label{tab:qwta_variant_comparison}% Original Table II label retained to avoid broken references
	\squeezetable
	\setlength{\tabcolsep}{2.2pt}% Seven columns; 9 pt table text with compact column spacing
	\small
	\begin{ruledtabular}
		\begin{tabular}{ccccccc}
			& \multicolumn{4}{c}{Classical baselines}
			& \multicolumn{2}{c}{QWTA models} \\
			\cline{2-5} \cline{6-7}
			MASK & GAT & GCN & Transformer & GIN
			& QWTA-Base & QWTA-GR \\
			\hline

			0
			& $0.10320 \pm 0.00057$
			& $0.10123 \pm 0.00032$
			& $\underline{0.10100 \pm 0.00122}$
			& $0.16410 \pm 0.08726$
			& $0.09994 \pm 0.00037$
			& $\mathbf{0.09755 \pm 0.00007}$ \\

			1
			& $0.10750 \pm 0.00028$
			& $\underline{0.10643 \pm 0.00101}$
			& $0.11020 \pm 0.00056$
			& $0.10755 \pm 0.00120$
			& $0.10736 \pm 0.00091$
			& $\mathbf{0.10595 \pm 0.00134}$ \\

			2
			& $0.11745 \pm 0.00276$
			& $\underline{\mathbf{0.11465 \pm 0.00055}}$
			& $0.11700 \pm 0.00014$
			& $0.11680 \pm 0.00184$
			& $0.11496 \pm 0.00375$
			& $0.11625 \pm 0.00007$ \\

			3
			& $0.12380 \pm 0.00410$
			& $0.12220 \pm 0.00089$
			& $\underline{\mathbf{0.12150 \pm 0.00028}}$
			& $0.12200 \pm 0.00113$
			& $0.12383 \pm 0.00237$
			& $0.12387 \pm 0.00078$ \\

			4
			& $0.13110 \pm 0.00453$
			& $\underline{0.12300 \pm 0.00070}$
			& $\underline{0.12300 \pm 0.00000}$
			& $0.12575 \pm 0.00092$
			& $0.12374 \pm 0.00099$
			& $\mathbf{0.12130 \pm 0.00410}$ \\

			5
			& $0.14535 \pm 0.00233$
			& $\underline{\mathbf{0.13913 \pm 0.00061}}$
			& $0.14190 \pm 0.00127$
			& $0.14260 \pm 0.00156$
			& $0.14042 \pm 0.00180$
			& $0.14220 \pm 0.00170$ \\

			6
			& $0.15780 \pm 0.00438$
			& $0.15120 \pm 0.00061$
			& $\underline{0.15000 \pm 0.00014}$
			& $0.15445 \pm 0.00276$
			& $\mathbf{0.14844 \pm 0.00173}$
			& $0.15430 \pm 0.00042$ \\

			7
			& $0.17430 \pm 0.00339$
			& $0.17190 \pm 0.00131$
			& $\underline{0.16955 \pm 0.00177}$
			& $0.17195 \pm 0.00021$
			& $0.17077 \pm 0.00514$
			& $\mathbf{0.16595 \pm 0.00007}$ \\

			8
			& $0.19305 \pm 0.00106$
			& $0.18690 \pm 0.00099$
			& $0.18740 \pm 0.00014$
			& $\underline{0.18685 \pm 0.00021}$
			& $0.19067 \pm 0.00146$
			& $\mathbf{0.18457 \pm 0.00254}$ \\

			9
			& $0.22705 \pm 0.00092$
			& $\underline{\mathbf{0.22325 \pm 0.00007}}$
			& $0.22410 \pm 0.00147$
			& $0.22580 \pm 0.00269$
			&$0.22723 \pm 0.00335$
			& $0.22340 \pm 0.00057$ \\

			10
			& $0.27145 \pm 0.00049$
			& $\underline{0.26973 \pm 0.00040}$
			& $0.26990 \pm 0.00051$
			& $0.27235 \pm 0.00078$
			& $0.27902 \pm 0.00380$
			& $\mathbf{0.26885 \pm 0.00007}$ \\
			
			\hline

			average
			& $0.15928$
			& $\underline{0.15542}$
			& $0.15596$
			& $0.16275$
			& $0.15695$
			& $\mathbf{0.15493}$ \\
		\end{tabular}
	\end{ruledtabular}
\end{table*}

Table~\ref{tab:main_results} reports the validation MSE of four classic graph propagation baselines and two QWTA models under 11 historical missing settings ranging from MASK0 to MASK10. All models receive the same input features: normalized speed and the timestamps of retained observations. The table caption specifies how validation MSE is aggregated across the retained runs. Macro averages are computed before rounding the displayed per-MASK values. For comparison, we refer to the minimum MSE of the four classical baselines, namely the underlined values in the table, as Best Classical. We define
\begin{equation}
	\Delta= \text{MSE}_\text{QWTA} − \text{MSE}_\text{Best Classical}
\end{equation}
as the difference between the MSE obtained by the QWTA models and that of the classical baselines.

\begin{figure}[H]
 \centering
 \includegraphics[width=\linewidth]{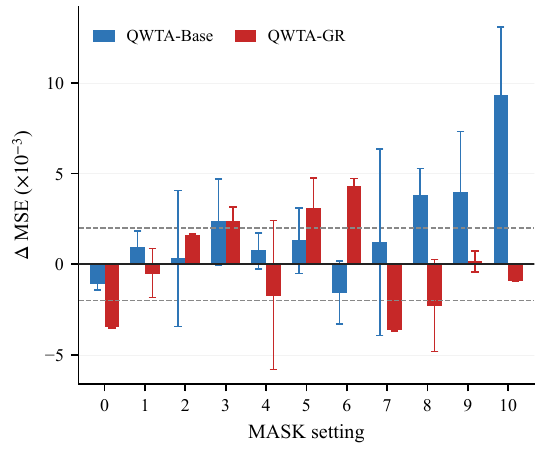}
 \caption{The validation MSE gaps between QWTA-Base/QWTA-GR and the best classical baseline for each MASK setting are computed from the model means reported in Table~\ref{tab:main_results}. The solid blue and red bars correspond to $\Delta_\text{Base}$ and $\Delta_\text{GR}$, respectively. The bar chart plots the mean of each model minus the fixed mean of the best classical baseline. Error bars represent the corresponding sample standard deviations of the models reported in Table~\ref{tab:main_results}. The dashed lines at $\pm0.002$ mark the descriptive approximate tie range.}
 \label{fig:delta_gr_best_classical}
\end{figure}

Figure~\ref{fig:delta_gr_best_classical} visualizes the
MSE gaps between QWTA-Base/QWTA-GR and Best Classical
for MASK0--MASK10. QWTA-Base achieves lower mean validation
MSE than Best Classical under MASK0 and MASK6, with
$\Delta_{\mathrm{Base}}=-0.00106$ and $-0.00156$, respectively. Using the descriptive threshold $|\Delta|<0.002$,
QWTA-Base is approximately tied with Best Classical under
seven settings (MASK0, MASK1, MASK2, MASK4, MASK5, MASK6
and MASK7). It performs worse under the remaining four
settings (MASK3, MASK8, MASK9 and MASK10), with gaps ranging
from $0.00233$ to $0.00929$. Its macro-averaged validation MSE is $0.15695$, compared
with $0.15542$ for GCN, the best single classical baseline.
Its macro-averaged gap relative to Best Classical is
approximately $0.00194$. These results indicate that exact CTQW spectral propagation
is competitive under several missing settings but does
not outperform the classical references on average. This motivates examining phase soft truncation and gated
residual fusion as extensions to the base model.

With the above mechanism introduced, the macro-averaged validation MSE of QWTA-GR drops to 0.15493, which is the lowest among all models. It outperforms GCN, the best standalone classical baseline, by approximately 0.00049 (a relative reduction of around 0.32$\%$), and is marginally lower than the macro-averaged Best Classical across all MASK settings of 0.15501, with a difference of −0.00008 and a relative reduction of roughly 0.05$\%$. It should be emphasized that Best Classical serves as a strict reference constructed by selecting the optimal result from the four baselines a posteriori for each setting.

In terms of per-setting results, QWTA-GR achieves the lowest mean validation MSE among all models under six of the eleven configurations (MASK0, MASK1, MASK4, MASK7, MASK8 and MASK10).QWTA-Base achieves the lowest mean under MASK6, while classical baselines achieve the lowest means under the remaining four configurations (MASK2, MASK3, MASK5 and MASK9).To avoid overinterpreting tiny numerical differences, we mark cases with \(|\Delta|<0.002\) as approximately tied. Further classification based on the threshold \(|\Delta_{\text{GR}}|<0.002\) yields the following observations: QWTA-GR substantially outperforms Best Classical under three settings (MASK0, MASK7, MASK8) with error reductions ranging from 0.00228 to 0.00360, and performs notably worse under three settings (MASK3, MASK5, MASK6) with performance gaps between 0.00237 and 0.00430. The remaining five settings (MASK1, MASK2, MASK4, MASK9, MASK10) fall within the approximately tied range. Consequently, the performance improvement of QWTA-GR exhibits clear mask dependence. Its advantages hold only for certain specific missing patterns instead of being consistent across all missing scenarios.

It is worth noting that MASK0, one of the settings with the largest error reduction, corresponds to fully regular observations, where all adjacent valid intervals \(\Delta t_k\) are equal. Under this condition, the phase encoding mechanism for irregular intervals does not take effect. The performance gain under this setting should mainly stem from the complex valued spectral propagation and stabilization structure itself. The independent contribution of interval encoding will be separately examined by the controlled variant QWTA FixedTime in Sec.~\ref{ssec:ablation}.

Overall, the results of QWTA-Base establish the baseline capability of pure CTQW propagation. QWTA-GR demonstrates that after incorporating phase soft truncation and gated residual fusion on top of the base model, the model has a slightly lower macro average validation MSE than the per-MASK classical reference over the controlled settings. It achieves the best overall performance in six configurations. This conclusion is limited to the validation set. The mask dependence of performance tradeoffs and its correlation with the spectral structure of traffic signals and learned phase scales will be further analyzed in subsequent subsections.

\subsection{Experimental analysis of the time scale parameter and phase soft clipping}
\label{ssec:fidelity}

\begin{figure}[t]
	\centering
	\includegraphics[width=\linewidth]{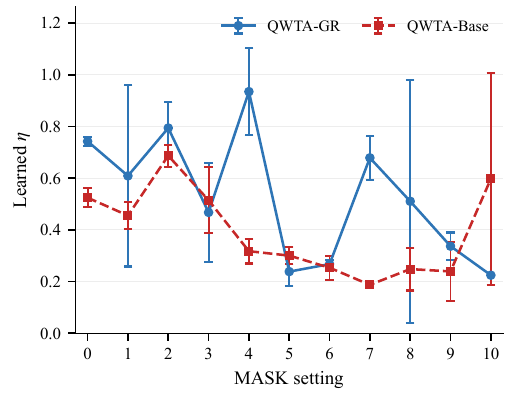}
	\caption{
		Learned signed time scale $\eta$ at the validation-best
checkpoints for MASK0–MASK10. Both curves show the
mean and sample standard deviation within each MASK.
	}
	\label{fig:learned_eta_by_mask}
\end{figure}

We first analyze the learned time scale parameter \(\eta\) of the two models under different MASK settings. In Fig.~\ref{fig:learned_eta_by_mask} we compare the signed shared \(\eta\) obtained from validation-best checkpoints instead of mixing best and last training states. Within each MASK setting, we report sample standard deviations.

The individual \(\eta\) values of QWTA-Base span 0.1240 to 0.8447. Its per-MASK means vary from 0.1875 under MASK7 to 0.6873 under MASK2. MASK10 yields a mean of \(0.5978\pm0.4104\), with individual seed realizations equal to 0.1240, 0.8245 and 0.8447. This large dispersion indicates that MASK10 does not produce a consistently high learned phase scale. The retained checkpoint values of QWTA-GR range from 0.1788 to 1.0541, and its per-MASK means lie between 0.2249 and 0.9351. The error bars for QWTA-GR characterize run-to-run dispersion across the available trials. These error bars reflect differences in both configuration and seed. Neither sequence exhibits a monotonic trend with the number of valid observations. The plotted curves illustrate mask-dependent learned phase scales, but they do not provide a causal account for the observed discrepancies in forecasting error.

Next, to explain the differing sensitivity of phase soft clipping to various inputs, we investigate the spectral energy distribution of the input graph signals under the Laplacian eigenbasis. For the spectral analysis of the raw traffic data, we use the normalized Laplacian of the graph with self-loops. Following Eq.~\eqref{eq:laplacian_eigendecomp}, we write its spectral decomposition as $L_{\mathrm{raw}}=V_{\mathrm{raw}}\Lambda_{\mathrm{raw}}V_{\mathrm{raw}}^\top$, with eigenpairs $(\lambda_n^{\mathrm{raw}},\mathbf v_n^{\mathrm{raw}})$. The METR-LA graph has \(N=207\) nodes, and its reported spectral range satisfies \(\lambda_n^{\mathrm{raw}}\in[0,1.2076]\), with an upper bound below that of the conventional spectral interval \([0,2]\) for the normalized Laplacian operator.

For QWTA-GR, the soft phase truncation reduces local phase sensitivity at large unclipped phase magnitudes and introduces a deviation from the unclipped phase. To quantify such discrepancies, we compute their difference and perform a Taylor expansion, that is
\begin{equation}
	\varphi_{nk}
	-
	\tilde{\varphi}_{nk}
	=
	\frac{\varphi_{nk}^{3}}{3\pi^2}
	+
	O(\varphi_{nk}^{5}).
\end{equation}
In the small-phase approximation, the leading term of the phase deviation is proportional to the cube of the spectral eigenvalue, so the second-order expansion of fidelity loss involves sixth-order spectral weighting. Following the signal spectral energy weighting in Ref.~\cite{Anis2014SpectralMoments}, we define the sixth spectral moment as
\begin{equation}
	\mu_6(\mathbf{x})
	=
	\sum_{n=1}^{N}
	\left|(\mathbf v_n^{\mathrm{raw}})^\top\mathbf x\right|^2(\lambda_n^{\mathrm{raw}})^6.
	\label{eq:mu6}
\end{equation}
Here, $\left|(\mathbf v_n^{\mathrm{raw}})^\top\mathbf x\right|^2$ is the energy weight of spectral mode $n$. This moment helps compare the sensitivity of different inputs at the same graph spectrum and effective phase scale.

In our experiments, the sixth-order spectral moment of the temporal mean traffic speed signal is \(\mu_6(\overline{\mathbf{x}}) = 0.00879\). For 100 randomly sampled traffic speed signals, we obtain \(\overline{\mu}_6^{\mathrm{traffic}} = 0.02318 \pm 0.01213\), while 200 unstructured random inputs in the node domain yield \(\overline{\mu}_6^{\mathrm{random}} = 0.17243 \pm 0.01966\). The average sixth-order spectral moment of random inputs is approximately 7.44 times that of real traffic inputs, demonstrating that the spectral energy of traffic speed signals concentrates more on low-frequency Laplacian eigenmodes.

To evaluate the propagation step in Algorithm~\ref{alg:qwta_gr_forward}, take the graph signal $\mathbf c_{b,\ell}^{(k)}=\mathbf C^{(k)}[b,:,\ell]$ from the step-specific convolution output, where $b\in\{1,\ldots,B\}$ is the sample index and $\ell\in\{1,\ldots,d\}$ is the hidden-channel index. Apply the untruncated and soft-clipped operators to the same normalized signal with the same graph spectrum, $\Delta t_k$, and $\eta$. Compare the two outputs before gated residual fusion.
For each nonzero signal $\mathbf c_{b,\ell}^{(k)}$, define its unit-norm version
$\widehat{\mathbf c}_{b,\ell}^{(k)}=\mathbf c_{b,\ell}^{(k)}/\|\mathbf c_{b,\ell}^{(k)}\|_2$.
Its propagation fidelity is

\begin{equation}
	F(\widehat{\mathbf c}_{b,\ell}^{(k)},\Delta t_k)
	=
	\left|
	\left(
	U(\Delta t_k)\widehat{\mathbf c}_{b,\ell}^{(k)}
	\right)^{\dagger}
	\left(
	\widetilde U(\Delta t_k)\widehat{\mathbf c}_{b,\ell}^{(k)}
	\right)
	\right|^2.
	\label{eq:hin_fidelity}
\end{equation}

We ran one saved QWTA-GR checkpoint per MASK with the time intervals in Algorithm 2, including $\Delta t_K=12-t_K$. We used the same 128 randomly selected test windows for each MASK and sampled 8 hidden channels without replacement at each retained step. This yielded 78,848 correlated post-convolution observations, each over 207 nodes.

When $F(\widehat{\mathbf c}_{b,\ell}^{(k)},\Delta t_k)$ approaches 1, the outputs obtained from untruncated propagation and soft phase truncation propagation given the same hidden input are highly consistent. A noticeable drop in $F(\widehat{\mathbf c}_{b,\ell}^{(k)},\Delta t_k)$ indicates that soft truncation exerts a strong modulation on the spectral propagation result of this hidden channel. With the graph spectrum \(\lambda_n\) fixed, the strength of soft truncation is mainly governed by the effective phase scale \(s = |\Delta t_k \eta|.\) Across all sampled MASK settings, the maximum observed $s$ is approximately 6.7425.

\begin{figure}[t]
	\centering
	\includegraphics[width=\linewidth]{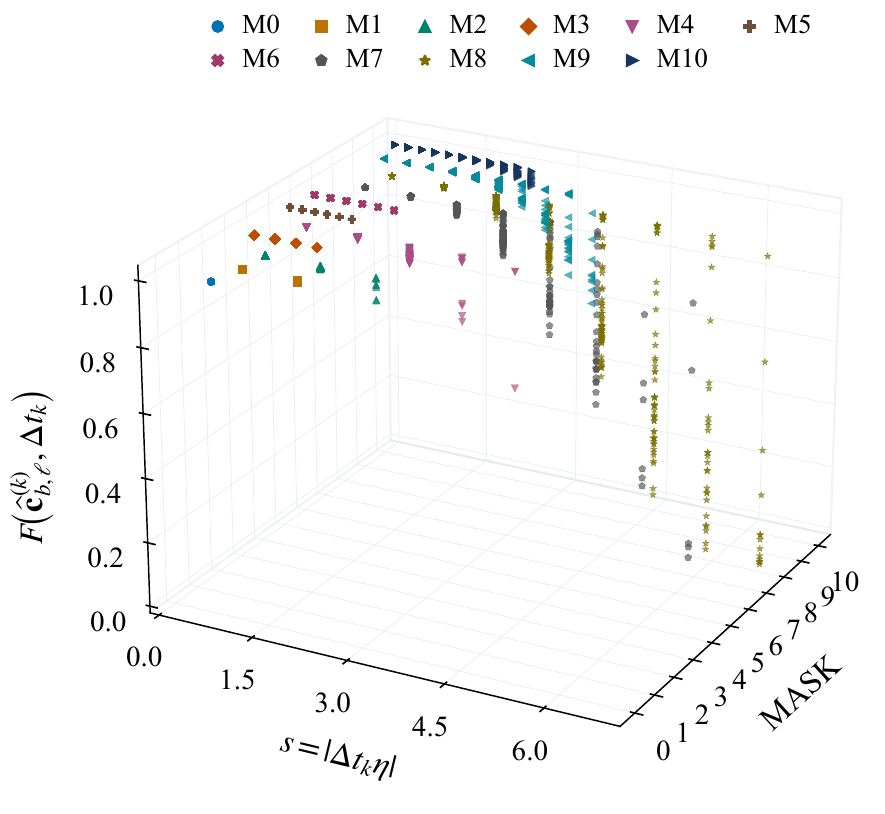}
	\caption{
		Fidelity between untruncated and soft-clipped propagation on the same post-convolution QWTA-GR signals across MASK settings. Each point represents one normalized nonzero channel from $\mathbf C^{(k)}$. The two outputs are compared before gated residual fusion. At large $s$, spectral phase compression can reduce fidelity.
	}
	\label{fig:fig_hin_raw_s_vs_F}
\end{figure}

Figure~\ref{fig:fig_hin_raw_s_vs_F} displays 8,911 observations, selected with a fixed random seed: up to 800 per MASK, supplemented by the records with minimum and maximum fidelity at each distinct $s$ within each MASK. Each point represents one post-convolution hidden channel graph signal. The MASK axis numbers 0--10 identify categories; the legend above the plot maps colors and shapes to these categories. Layer spacing has no physical meaning. Because the sample includes extrema and points may overlap, display density does not estimate frequency.

The following statistics are pooled over all extracted observations. Overall, 95.06\% of observations have $F(\widehat{\mathbf c}_{b,\ell}^{(k)},\Delta t_k)\geq0.99$, and 1.51\% have $F(\widehat{\mathbf c}_{b,\ell}^{(k)},\Delta t_k)<0.9$. Observation counts differ across MASK settings because the number of retained steps varies. Under MASK8, 17.38\% have $F(\widehat{\mathbf c}_{b,\ell}^{(k)},\Delta t_k)<0.9$; the median is $0.99669$, the 10th percentile is $0.66872$, and the minimum is $0.00239$.

At small $s$, the clipped and unclipped phases are close. At larger $s$, mode-dependent phase compression can cause cancellation in their spectral overlap and reduce $F(\widehat{\mathbf c}_{b,\ell}^{(k)},\Delta t_k)$. Channels at similar $s$ can respond differently because their spectral energy weights differ; $F(\widehat{\mathbf c}_{b,\ell}^{(k)},\Delta t_k)$ need not decrease monotonically with $s$. Both operators preserve the input norm, so lower fidelity describes a change in relative phase and state direction.

Figure~\ref{fig:fig_hin_raw_s_vs_F} shows that most post-convolution signals retain high propagation fidelity under soft phase clipping. Lower fidelity values occur mainly at larger phase scales $s$, with the most pronounced low-fidelity tail in MASK8.

\subsection{Module ablation experiment}
\label{ssec:ablation}

\begin{figure}[t]
	\centering
	\includegraphics[width=\linewidth]{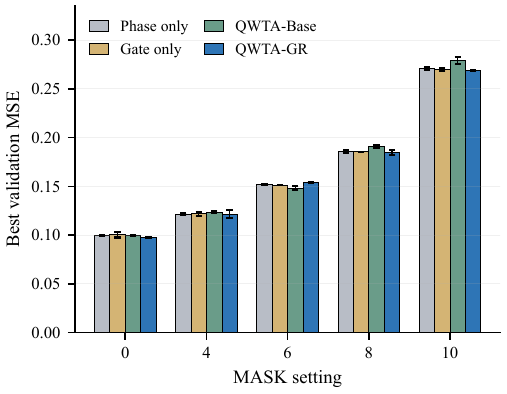}
	\caption{
		Component ablation results for phase modulation and gated residual modules.
		The figure compares the Phase-only, Gate-only, QWTA-Base, and full QWTA-GR variants under five representative MASK settings in terms of validation MSE.
		Phase-only retains phase soft clipping and replaces learned residual weights with a fixed sum; Gate-only retains learned residual weights and removes phase soft clipping. Both retain the QWTA-GR temporal gate and the same post-convolution residual input.
		QWTA-Base denotes the basic CTQW spectral propagation backbone without phase soft clipping or gated residual fusion.
		QWTA-GR denotes the full model with both stabilization mechanisms.
	}
	\label{fig:ablation_component_bar}
\end{figure}

To assess the contributions of phase soft clipping and gated residual weighting within QWTA-GR, we conduct two component ablations derived from the full QWTA-GR model. Both ablations retain the QWTA-GR temporal gate, convolution, residual input source and output processing; only the targeted component is changed. As illustrated in Fig.~\ref{fig:ablation_component_bar}, under four representative missing settings including MASK0, MASK4, MASK8 and MASK10, QWTA-GR achieves lower validation MSE than QWTA-Base, with error reductions of $0.00239$, $0.00244$, $0.00610$ and $0.01017$, respectively. Across all five representative settings shown in
Fig.~\ref{fig:ablation_component_bar} (MASK0, MASK4, MASK6, MASK8 and MASK10), the macro-averaged validation MSE decreases from $0.16836$ for QWTA-Base to $0.16531$ for QWTA-GR. Under MASK6, QWTA-Base has the lower MSE: $0.14844$ versus $0.15430$ for QWTA-GR, a difference of $0.00586$.

We then compare the full QWTA-GR model with its two component ablations. The macro-averaged validation MSE values are $0.16596$ after replacing learned residual weights with a fixed sum (Phase-only) and $0.16584$ after removing phase soft clipping (Gate-only). Both values are close to $0.16531$ of the full QWTA-GR, yet discrepancies remain across specific missing patterns. QWTA-GR outperforms the better of the two single-mechanism variants under MASK0, MASK4, MASK8 and MASK10, supporting the combined use of phase soft clipping and gated residual weighting under these settings. Under MASK6, the Gate only variant yields slightly lower error than QWTA-GR, indicating that phase clipping does not reduce validation error under every MASK setting. Together, these comparisons support retaining both components in QWTA-GR based on the macro average validation MSE, while the MASK6 result shows that the benefit depends on the missing pattern.

Following the component ablations, we construct two propagation controls from the full QWTA-GR model. QWTA-Diffusion replaces the soft-clipped complex phase operator with heat diffusion, using the observed intervals and a positive diffusion rate learned jointly across all propagation steps. QWTA-FixedTime retains the QWTA-GR phase operator, including soft phase truncation and the trainable phase scale $\eta$, and sets every propagation interval, including the final one, to 1. Both variants retain the QWTA-GR input features, temporal fusion, gated residual fusion, and output layer.

\begin{figure}[t]
	\centering
	\includegraphics[width=\linewidth]{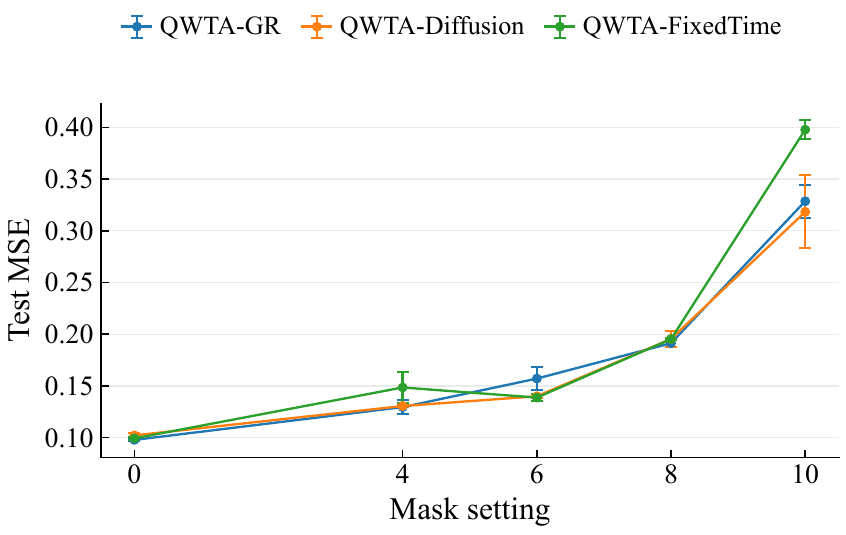}
	\caption{
		Test MSE comparison between QWTA-GR and two controlled propagation mechanism variants under different MASK settings.
		QWTA-GR uses CTQW-inspired complex phase spectral propagation parameterized by valid observation intervals.
		QWTA-Diffusion replaces unit-modulus complex phase propagation with classical diffusion-type spectral propagation.
		This variant tests the contribution of phase propagation relative to dissipative diffusion propagation.
		QWTA-FixedTime fixes the propagation time interval and tests the role of encoding nonuniform intervals.
		Except for the propagation operator and its temporal parameterization, the three models share the same input features, temporal fusion structure, gated residual structure and output prediction layer.
		Markers show the mean test MSE across two retained runs per model and MASK, and error bars show the sample standard deviation across those runs.
	}
	\label{fig:qwta_mechanism_mse}
\end{figure}

Fig.~\ref{fig:qwta_mechanism_mse} reports the test MSE of the three models under different MASK settings. QWTA-GR achieves lower test error than both QWTA-Diffusion and QWTA-FixedTime for MASK0, MASK4 and MASK8. In addition, QWTA-GR does not outperform the two controlled variants under MASK6, and it is inferior to QWTA-Diffusion under MASK10. At MASK0, QWTA-GR and FixedTime both use unit intervals and apply the same propagation kernel when the graph and phase scale are held fixed. The test errors in Fig.~\ref{fig:qwta_mechanism_mse} describe the performance of the trained variants under each MASK.

\section{Quantum circuit design}
\label{ssec:quantum_circuit}

To evaluate the computational cost of different models, we obtain parameter counts from the actual models and measure single batch forward inference time and peak CUDA memory consumption under identical input sizes. The complexity test adopts an input configuration with \(B = 16\), \(K = 12\), \(N = 207\), \(F = 2\), and hidden dimension \(d = 64\), corresponding to an input tensor of size \(\mathbb{R}^{16\times12\times207\times2}\). After 10 warm up runs, the average forward pass time is computed over 50 forward propagations. The current test uses \(N = 207\), matching the number of nodes in METR LA, to simulate a chain graph. Therefore, the parameter count directly reflects model size, while inference time and memory consumption are reported as measurements at the stated input scale.

Table~\ref{tab:model_complexity} presents the complexity comparison of different models under identical input scales. For $K=12$, $F=2$ and $d=64$, the algorithm-specified QWTA-Base and QWTA-GR architectures contain 349,763 and 357,978 real-valued trainable parameters, respectively. Both are comparable in size to Transformer with 352,066 parameters. The increase of 8,215 comprises 8,191 additional temporal gate parameters and 24 step-specific residual gate scalars.

It should be noted that the complexity results in Table~\ref{tab:model_complexity} correspond to the cost of numerically simulating the QWTA propagation layer on conventional GPUs. Within this implementation, CTQW type propagation requires explicit construction and application of the complex valued propagation operator \(U(\Delta t_k) \in \mathbb{C}^{B\times N\times N}\), which leads to high inference time and memory overhead. From the perspective of physical implementation, however, the spectral propagation layer of QWTA-Base essentially corresponds to the continuous time evolution \(e^{-iL\tau}\) induced by Hamiltonian $L$. QWTA-GR instead uses the soft-clipped spectral generator defined below. If this propagation process is directly realized using programmable quantum devices or photonic devices, the propagation operator can be provided by the natural evolution of the physical system. The classical side no longer needs to explicitly construct and store the full \(N\times N\) complex valued propagation matrix. Consequently, QWTA exhibits high complexity under classical simulation implementations, while its propagation layer holds the potential to reduce classical computational burden via quantum physical hardware.

\begin{table}[t]
	\caption{
		Complexity comparison of different models under the same input size.
		Time denotes the average forward propagation time for one batch.
		Memory denotes peak allocated CUDA memory in MiB.
	}
	\label{tab:model_complexity}
	\squeezetable
	\begin{ruledtabular}
		\begin{tabular}{cccc}
			Model & Params & Time / batch (ms) & Memory (MiB) \\
			\hline
			GCN & $200{,}002$ & $10.237$ & $70.54$ \\
			GIN & $249{,}922$ & $11.690$ & $70.73$ \\
			GAT & $201{,}538$ & $26.424$ & $70.55$ \\
			Transformer & $352{,}066$ & $27.100$ & $71.13$ \\
			QWTA-Base & $349{,}763$ & $19.732$ & $74.67$ \\
			QWTA-GR & $357{,}978$ & $15.486$ & $77.51$ \\
		\end{tabular}
	\end{ruledtabular}
\end{table}

\begin{figure}[t]
\centering
% Fig.8: vector artwork, final font sizes matched to Fig.6.
\includegraphics[width=\columnwidth]{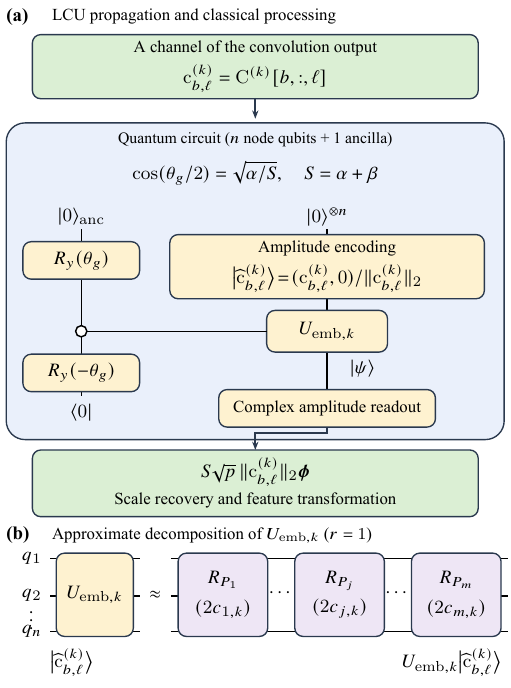}
\caption{
			Hybrid quantum-classical gate circuit for the QWTA-GR propagation layer and a schematic decomposition of the core propagation operator.
			(a) Overall computation flow of the QWTA-GR propagation layer. A channel \(\mathbf c_{b,\ell}^{(k)}\) of the convolution output \(\mathbf C^{(k)}\) is written into the node register by amplitude encoding. Ancilla qubits prepare LCU weights and control the execution of the embedded QWTA-GR propagation operator $U_{\mathrm{emb},k}$. Classical modules multiply the complex amplitudes obtained from readout by the scale factor \(S\sqrt p\,\|\mathbf c_{b,\ell}^{(k)}\|_{2}\) and then apply feature transformations to obtain real-valued features.
			(b) Conceptual gate-level decomposition of the propagation operator $U_{\mathrm{emb},k}$. The node register consists of $n=\lceil \log_2 N\rceil$ qubits, and the full propagation gate can be approximately implemented by a sequence of Pauli product rotation modules $R_{P_j}(2c_{j,k})$ (shown for \(r=1\)).
	}
	\label{fig:quantum_u_circuit}
\end{figure}

In this article, we present a rough proposal for implementing the QWTA-GR propagation layer via quantum gate circuits in this work. Notably, we do not fully quantize the entire spatiotemporal forecasting network. Instead, we focus on providing a gate level implementation for the core operator \(\widetilde U(\Delta t_k)\) and its gated residual fusion within the propagation layer. In this propagation subroutine, the convolution output \(\mathbf C^{(k)}\) is first encoded into node registers by channel. The quantum processor is mainly responsible for time dependent graph propagation and the LCU gated residual combination. For a nonzero channel \(\mathbf c_{b,\ell}^{(k)}=\mathbf C^{(k)}[b,:,\ell]\), we construct a layer propagation operation incorporating historical information, \(M_k=\alpha\widetilde U(\Delta t_k)+\beta I\), where $\alpha=\alpha^{(k)}$ and $\beta=\beta^{(k)}$ are the residual weights in Eq.~\eqref{eq:Gated_residual}. We write \(S=\alpha+\beta\). The LCU ancilla is prepared with amplitudes \(\sqrt{\alpha/S}\) and \(\sqrt{\beta/S}\), with propagation on branch \(0\) and identity on branch \(1\). Subsequent scaling recovery and complex feature transformations remain performed by classical modules (as illustrated in Fig.~\ref{fig:quantum_u_circuit}(a)).

We normalize the nonzero channel $\mathbf c_{b,\ell}^{(k)}$ and pad it with zeros to $D=2^{\lceil\log_2N\rceil}$ dimensions. The resulting amplitude-encoded state is denoted by $\bigl|\widehat{\mathbf c}_{b,\ell}^{(k)}\bigr\rangle$. Its amplitude vector in the computational basis is
\[
\frac{1}{\|\mathbf c_{b,\ell}^{(k)}\|_2}
\begin{bmatrix}
\mathbf c_{b,\ell}^{(k)}\\
\mathbf 0_{D-N}
\end{bmatrix}.
\]

After unpreparation, the probability of ancilla outcome \(0\) is \(p=\|M_k\mathbf c_{b,\ell}^{(k)}\|_{2}^2/(S^2\|\mathbf c_{b,\ell}^{(k)}\|_{2}^2)\). Conditioned on this outcome, let $|\psi\rangle$ denote the normalized state of the node register. Its complex amplitude vector on the active nodes is defined as
\begin{equation}
\boldsymbol{\phi}
=
\bigl(
\langle0|\psi\rangle,\ldots,
\langle N-1|\psi\rangle
\bigr)^\top
\in\mathbb C^N.
\label{eq:postselected_amplitudes}
\end{equation}

Once the complex amplitudes have been obtained with a phase convention consistent with the operator representation above, multiplying the normalized complex amplitude vector $\boldsymbol{\phi}$ by the scale factor $S\sqrt p\,\|\mathbf c_{b,\ell}^{(k)}\|_2$ recovers the propagation output $M_k\mathbf c_{b,\ell}^{(k)}$:
\begin{equation}
M_k\mathbf c_{b,\ell}^{(k)}
=
S\sqrt p\,\|\mathbf c_{b,\ell}^{(k)}\|_{2}\boldsymbol{\phi}.
\label{eq:classical_scale_recovery}
\end{equation}

Using the phases in Eq.~\eqref{eq:phase_clip}, we pad each input channel with zeros to \(D=2^n\) dimensions and write the corresponding embedded quantum evolution operator as
\begin{equation}
U_{\mathrm{emb},k}
=
\exp\!\left\{-i\left(
\left[V\operatorname{diag}(\tilde\varphi_{1k},\ldots,\tilde\varphi_{Nk})V^\top\right]\oplus 0_{D-N}
\right)\right\}.
\label{eq:embedded_propagator}
\end{equation}
Here $\oplus$ denotes a matrix direct sum, and \(0_{D-N}\) denotes a square zero block. The observation interval $\Delta t_k$ and time scale coefficient $\eta$ are already included in the spectral phases through Eq.~\eqref{eq:phase_clip}. For $N=207$ and $D=256$, the input has 49 added zero components, and evolution on the corresponding subspace is the identity.

The gate level implementation of the quantum evolution operator \(U_{\mathrm{emb},k}\) can follow the Pauli string exponentiation synthesis method widely adopted in digital quantum Hamiltonian simulation~\cite{u_trotter_pauli}. To this end, we expand the spectral phase matrix multiplied by $-i$ in the exponent as
\begin{equation}
	\left[V\operatorname{diag}(\tilde\varphi_{1k},\ldots,\tilde\varphi_{Nk})V^\top\right]\oplus 0_{D-N}
	=
	\sum_{j=1}^{m} c_{j,k} P_j,
	\label{eq:heff_pauli}
\end{equation}
where $P_j\in\{I,X,Y,Z\}^{\otimes n}$ is the $j$th Pauli word acting on the $n$-qubit node register. The coefficient $c_{j,k}\in\mathbb R$ is the corresponding expansion coefficient of the spectral phase matrix above, and $m$ is the total number of nonzero Pauli terms. The number of qubits for the node register is determined by the number of graph nodes, namely \(n = \lceil\log_2 N\rceil\). For instance, when the traffic graph contains \(N = 207\) sensor nodes, the node register requires \(n = 8\) qubits (\(D=256\)). The identity term contributes a phase to the controlled propagation branch.

Since different Pauli string terms generally do not commute exactly, product formulas can be adopted for their approximate decomposition in practical gate circuits

\begin{equation}
	U_{\mathrm{emb},k}
	\approx
	\left(
	\prod_{j=1}^{m}
	e^{-i c_{j,k} P_j/r}
	\right)^r ,
	\label{eq:u_trotter_pauli}
\end{equation}
where $r$ is the number of Trotter steps. When \(r = 1\), the above expression corresponds to the conceptual gate level expansion of the first order product formula. Increasing $r$ improves the approximation accuracy of the overall unitary evolution, yet also increases the number of quantum gates.

Eq.~\eqref{eq:u_trotter_pauli} indicates that the full propagation gate \(U_{\mathrm{emb},k}\) can be further decomposed into a sequence of Pauli string exponentiated terms. Each term \(e^{-ic_{j,k} P_j/r}\) can be implemented by a Pauli product rotation module, namely
\begin{equation}
	R_{P_j}\!\left(\frac{2c_{j,k}}{r}\right)
	=
	e^{-i c_{j,k} P_j/r}.
	\label{eq:pauli_rotation_block}
\end{equation}
In the schematic, we adopt the form with \(r = 1\) and denote this module as \(R_{P_j}(2c_{j,k})\). Here, \(P_j\) specifies the Pauli direction and the involved qubits acted on by this module, while \(c_{j,k}\) sets the corresponding rotation strength. For nonidentity Pauli words, the standard gate circuit structure of this module generally consists of basis transformation, a CNOT parity check network, $R_Z$ phase rotation and the corresponding inverse operations (Fig.~\ref{fig:quantum_u_circuit}(b)). Specifically, basis transformation maps $X$- or $Y$-type Pauli factors in \(P_j\) to the $Z$ basis. The CNOT network aggregates the parity phase over multiple qubits. An $R_Z(2c_{j,k}/r)$ gate is then applied on the target qubit. Finally, the original basis is restored via the inverse CNOT network and inverse basis transformation. Accordingly, \(U_{\mathrm{emb},k}\) can be represented not only as a global unitary propagation gate, but also interpreted as a sequence of Pauli rotation subcircuits acting on the node register.

For the soft truncated propagation operator of QWTA-GR, the parameter \(\eta\) and gated residual coefficients are fixed after training. Therefore, for a given valid time interval \(\Delta t_k\), $U_{\mathrm{emb},k}$ can be treated as a fixed unitary propagation gate. From this perspective, it can be compiled at gate level or numerically decomposed just like ordinary fixed unitary gates.

The full QWTA network still includes classical neural network components such as input projection, gated fusion, ComplexLinear, ModReLU, BatchNorm, ReLU and the final regression output layer. Figure~\ref{fig:quantum_u_circuit} shows eight node qubits and one LCU ancilla for \(N=207\). End-to-end resource accounting includes channel state preparation, controlled propagation at the required accuracy, postselection and repetitions, and phase-referenced complex amplitude readout.

\section{Discussion and Conclusion}
\label{sec:discussion_conclusion}

In this work, we propose a temporal graph forecasting framework QWTA based on CTQW for traffic graph signal prediction under irregular historical observations. The framework adopts the normalized Laplacian as the graph Hamiltonian and incorporates the propagation intervals $\Delta t_k$ into the phase modulation of the CTQW-type operator. Unlike classical graph neural networks that treat timestamps or time intervals as auxiliary input features, QWTA leverages valid observation intervals directly inside the propagation operator. It therefore provides a physically motivated time dependent graph propagation formulation for temporal graph modeling with missing historical observations.

Under the controlled historical missing settings MASK0–MASK10 on the METR-LA dataset, we validate the capacity of the QWTA framework from multiple perspectives including model design, controlled missing experiments and complexity analysis. Experimental results support the modeling strategy that treats nonuniform valid observation intervals as the time parameters of the CTQW inspired spectral propagation operator, and demonstrate that QWTA-GR achieves overall competitive prediction performance on the validation set for traffic forecasting with controlled missing historical observations. Meanwhile, QWTA propagation corresponds to CTQW Hamiltonian evolution. This offers a physically grounded route for further developing irregular temporal graph propagation on quantum processors or photonic devices.

	\section*{Acknowledgments}
	
	This research did not receive specific grants from funding
	agencies in the public, commercial or nonprofit sectors.
	
	\bibliographystyle{apsrev4-2}
	\bibliography{QWTA_cite}
	
\end{document}